\documentstyle[11pt,epsfig]{article}
\newcommand \be  {\begin{equation}}
\newcommand \bea {\begin{eqnarray} \nonumber }
\newcommand \ee  {\end{equation}}
\newcommand \eea {\end{eqnarray}}

 \def\(({\left(}
 \def\)){\right)}
\def\[[{\left[}
\def\]]{\right]}
\def\bi{\bibitem}

\def\ni{\noindent}
\def\ov{\over}

\def\eps{\epsilon}

\def \v{ {\vec v}}
\def \vx{\vec x}
\def \vy{\vec y}

\def \vr{\vec r}
\def \vq{\vec q}
\def \vf{\vec f}

\def \G1{ {G_{1}(\omega)}}

\def \e2{ {\eta(-\vec k,-\omega)}}

 \def\(({\left(}
 \def\)){\right)}
\def\[[{\left[}
\def\]]{\right]}
\def\bi{\bibitem}

\begin{document}

\title {Disordered systems and Burgers' turbulence}

\author{Marc M\'ezard}
  \date{\it Laboratoire de Physique Th\'eorique de l'Ecole Normale Sup\'{e}rieure \\
 (Unit\'e propre du CNRS,  associ\'ee
 \`a\ l'ENS et \`a\ l'Universit\'e de Paris Sud)\\ 
24 rue Lhomond, 75231 Paris Cedex 05, FRANCE\\
 email: \texttt{mezard@physique.ens.fr}.}

\maketitle

\abstract{
The problem of fully developped turbulence is to characterize
the statistical properties of the velocity field of a  stirred fluid described by
Navier stokes equations. The simplest scaling approach, due to Kolmogorov in
1941, gives a reasonable starting point, but it must be corrected
due to the failure of naive scaling giving  `intermittency' corrections
which are presumably associated with the existence of large
scale structures.
These scaling and intermittency properties can be studied analytically for
the case of stirred Burgers turbulence, a kind of simplified version of
Navier Stokes equations.
We use the mapping between Burgers' equation  and the problem of a directed
polymer in a random medium in order to study the fully developped
turbulence in the $d$ dimensional forced Burgers' equation, in
the limit of large dimensions. The
stirring force corresponds to a quenched (spatio temporal) random
potential for the polymer.
A replica symmetry breaking solution of the polymer problem provides the
full probability distribution of the velocity difference $u(r)$ between points
separated by a distance $r$ much smaller than the correlation length of the forcing.
This exhibits a very strong intermittency which is related to regions of shock
waves, in the fluid, and to the existence of metastable states
 in the directed polymer problem.
We also mention some recent
 computations on the finite dimensional problem,
based  on various analytical approaches
(instantons, operator product expansion, mapping to directed polymers),
as well as a conjecture on the relevance of Burgers equation
(with the length scale playing the role of time) for
the description of the functional renormalisation group
flow for the effective pinning potential of a manifold pinned by impurities.
Preprint LPTENS 97/66.
}

\section{Introduction}

The understanding of fully developped turbulence is a well posed problem
of mathematical physics \cite{turbu}. It has been around for more than fifty years but
 in spite of many interesting developments very little is known for sure.
In such a situation, it is clear that this field would
 benefit a lot from  the existence of a solvable model which, even if not realistic,
would exhibit the supposed properties of turbulence such
as scaling and intermittency, and allow for a detailed study of them.
The same role was played in the field of phase transitions by the Ising model and
its solution by Onsager, hence the quest for an "Ising model of turbulence". Two
candidates which have received a lot of attention recently are the stirred Burgers equation
on the one hand, and the diffusion of a passive scalar in a random -gaussian-
velocity field \cite{pas_scal} on the other hand.
 I shall describe the former, focusing onto
the solution of the
infinite dimensional problem  which we proposed with J.P. Bouchaud and G.Parisi
\cite{BMP}.
Burgers turbulence displays very interesting relationship with a fundamental problem
 of the statistical
mechanics of disordered systems, that of directed polymers in random media. 
Motivated both by the striking convergence between these two important fields and by 
my own background on disordered systems, I shall put particular emphasis on this relationship,
 which has already been quite useful so far, and will
probably lead to other important developments in the next years.

The next section is a rapid presentation of the problem of fully developped turbulence.
 Sect.3 describes Burgers turbulence and its mapping onto a problem of
directed polymers. Sect.4 contains a sketch of the solution in large dimensions.
Sect.5 presents briefly some of the issues in the finite dimensional problem.

\section{Fully developped turbulence}
This short description is included here to set the stage for the next sections. It
is very sketchy, I refer the reader to recent books on the subject for more
precise presentations \cite{turbu}.

Consider a three-dimensional fluid, the velocity of which verifies Navier-Stokes equations,
 in the incompressible
limit:
\be
{\partial \v \over \partial t} + (\v \cdot \vec \nabla) \ \v
= \nu \nabla^2 \v + \vec f(\vec x,t) - \vec \nabla p
\ \ ; \ \ 
\vec \nabla .\v=0
\label{NS}
\ee
The coefficient $\nu$ is the viscosity. The force $\vec f$ is the
external stirring force, which is supposed to inject energy into the system on a length scale $\Delta$. Specifically, one can take for instance 
a gaussian distributed random force
\footnote{
A different  problem, related to deterministic chaos, concerns how
a non random force becomes equivalent to a random one. I do not address
this issue  here.
}
 characterized by its two moments:
\be
<f^\mu(\vx,t)>=0
\ \ ; \ \
<f^\mu(\vx,t) f^\nu(\vy,t')>=\epsilon \delta(t-t') C^{\mu \nu}(|\vx-\vy|)
\ee
where the correlation function $C^{\mu \nu}(r)$ is normalised to one at the origin, and
decays rapidly enough when $r$ becomes larger or equal to $\Delta$ (I shall
not enter here into the details of the tensorial structure). The
parameter $\epsilon$ measures the energy injected into the fluid per
unit time and unit volume.

The problem is to understand the statistical properties of the velocity field
which is the solution of (\ref{NS}), in a regime which is both 
stationnary (i.e. neglecting initial transients) and isotropic
(we assume periodic boundary conditions), in the limit of
strong forcing. Ideally one might want to compute all moments
such as $<v(\vec x_1,t_1) \dots v(\vec x_q,t_q)>$, where the symbol 
$<.>$ means an average over various realizations of the random force.
Here I shall focus for simplicity onto the moments of the velocity
 difference between two points, at equal times:
\be
M_p=<|\v(\vec x+\vec r,t)-\v(\vec x,t)|^p>
\label{mom}
\ee
or equivalently the probability distribution function (pdf) of
$u=|\v(\vec x+\vec r,t)-\v(\vec x,t)|$.

This problem depends on three dimensionfull parameters, the viscosity $\nu$
with dimension ${length^2 \over time}$, the injection length
scale $\delta$, and the power injected per unit volume, $\epsilon$,
 with dimension ${length^2 \over time^3}$. From these one can build one adimensional number, the Reynolds number: $Re=({\epsilon \Delta^4 \over \nu^3})^{1/3}$.
By "strong forcing" is meant the limit of large $Re$. Another important length scale
is the dissipation length $l_d$ which is the length below which the viscosity
term becomes relevant: $l_d=\Delta/Re^{3/4}$.

The simplest description of what might happen in this problem 
is a scaling picture invented by Kolmogorov in 1941, usually called the
"K41 theory" \cite{Kol}. The energy is injected on a length scale $\Delta$ and dissipated
on the much smaller length scales of order $l_d$. In between lies the so called inertial 
regime where the energy is transferred towards smaller and smaller length scales
(a common metaphore is that of large eddies decaying into smaller ones). This
is called the inertial regime and is supposed to be universal (independent on the
details of the forcing for instance). The basic K41 hypothesis is that in 
this inertial regime $l_d<<r<<\Delta$, the velocity field
scales with respect to the distance:
\be
u=|\v(\vec x+\vec r,t)-\v(\vec x,t)| \sim r^\alpha
\label {scaling}
\ee
This scaling, denoted here with $\sim$, should be understood in law, meaning that 
the pdf of $u$ depends on the distance $r$ in the form $P(u)={1 \over r^\alpha}
\hat P({u \over r^\alpha})$, where the exponent $\alpha$ and the scaling
function $\hat P$ are to be determined. Scaling of other moments of $v$ assume 
a similar form, which basically amounts to saying that the instantaneous
properties of the flow in the inertial regime are {\it statistically 
invariant} under a simultaneous rescaling of the lengths by a factor $b$
and the velocities by the factor $b^\alpha$.
 
One can guess the value of $\alpha$ through a dimensional argument which
involves the following two hypotheses:

\noindent
- Locality in scale space. The statistics of $u$ depends on the separation
$r$ between the points, not on other length scales: $u \sim g(r,\nu,\eps)$.

\noindent
- Existence of the zero viscosity limit. The statistics of $u$ is well
defined when $\nu \to 0$. Therefore the pdf of $u$ depends only on $r$ and
$\eps$, which implies, dimensionally: $u \sim (\eps r)^{1/3}$, that is $\alpha=1/3$.

\begin{figure}
\centerline{\hbox{
\epsfig{figure=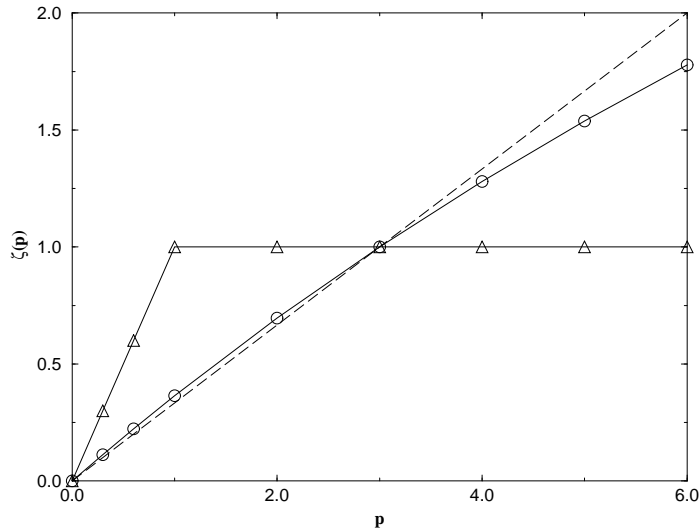,width=8cm,angle=-90}}
}
\caption{ Sketch of the `multifractal' spectrum $\zeta(p)$, giving the $r$ dependence of 
the $p$ th moment of the velocity field, both for the forced Burgers' equation (triangles) 
and hydrodynamical turbulence (circles). The dashed line is the K41 prediction.
}
\label{zetadep}
\end{figure}

This K41 scaling, derived from two simple and reasonable assumptions, implies
that the velocity difference between two points at distance $r$ behaves
as $r^{1/3}$ (Of course this holds only in the inertial range $\Delta >> r >>l_d$).
Going to Fourier space, it is easy to show that the total kinetic  
energy contained in the Fourier modes with wave vectors $\vq$ in
the shell $|\vq| \in [k,k+dk]$ scales as $k^{-5/3} dk$. Experimentally 
this scaling is very well verified at the level of the two point functions 
 (i.e. $M_2$) \cite{exp}. Analytically one of the few sure results concerns the
third moment of the $u$ pdf, which is known to scale linearly \cite{M3}:
$<u^3> = C r$. On the other hand, a careful study of the higher order
moments \cite{highmom}
favours a behaviour like  $<u^p> = C_p r^{\zeta_p}$, where the exponent 
$\zeta_p$ is smaller than the value $p/3$ which would be obtained by K41, 
indicating a failure of simple scaling (see Fig. \ref{zetadep}). 
This phenomenon has been given the name intermittency, because it is associated with 
the fact that the signal (e.g. the field $u$) has an intermittent structure 
both in space and in time, with bursts of activity separated by long quiet 
regions. This intermittency is related to the existence
of large scale structure in the flow, such as vorticity filaments \cite{fil}.
It suggests that the scaling is not uniform in space, but varies from point to point, 
a situation which can be described by a "multifractal" structure \cite{multifractal}.
All these observations have prompted many interesting developments the
description of which goes much beyond this presentation (see \cite{turbu}).

As we see the situation is somewhat confusing. The most commonly accepted 
conjecture is that some scaling properties hold in the inertial regime, such
as $<u^p> = C_p r^{\zeta_p}$, but that simple scaling does not hold:
$\zeta_p \ne \alpha p$. If true, this is an interesting situation, and
deriving these properties from the original stochastic partial differential equation
(\ref{NS}) is a major challenge of theoretical physics. In recent years
much attention has been paid to other "similar" but simpler problems,
providing an existence proof for intermittency together with some
hints on its origin.

\section{Burgers turbulence and directed polymers}
The Burgers equation is defined in any dimension $d$ (although most often it
is studied in $d=1$). The evolution equation for the velocity field is:
\be
{\partial \v \over \partial t} + (\v \cdot \vec \nabla) \ \v
= \nu \nabla^2 \v + \vec f(\vec x,t)
\ \ ; \ \ 
\v=- \vec \nabla h
\label{bueq}
\ee
It is a nonlinear stochastic partial differential equation which looks similar
to the Navier Stokes one (\ref{NS}). The only, but crucial, difference is
the second constraint: instead of an incompressible velocity, we consider the case where the
velocity is a gradient field. For consistency the forcing is also
supposed to be a gradient: 
\be
\vec f = - \vec \nabla \phi
\ee
Apart from this, the problem is defined as before: $\phi$ is supposed to
be a gaussian process, white noise in time, with a certain spatial
correlation length $\Delta$, and we want to study the statistical properties
of the flow at large $Re$. Because of the irrotational nature of
Burgers flow the large scale excitations are very different from
those of Navier Stokes. One should not consider this Burgers turbulence
in any sense as an approximation to real turbulence, but as a different
problem which we use as a laboratory for developping and testing
theoretical ideas. (Another difference is the non conservation
of energy by Burgers flow in $d>1$; Surprisingly this does not seem to
be crucial).

Clearly the scaling arguments of the previous section apply here as well,
and the simple K41 scaling would predict again that
 $u=|\v(\vec x+\vec r,t)-\v(\vec x,t)| \sim r^{1/3}$. As we shall see this is
too simple and there are strong intermittency effects.

\begin{figure}
\centerline{\epsfxsize=9cm
\epsffile{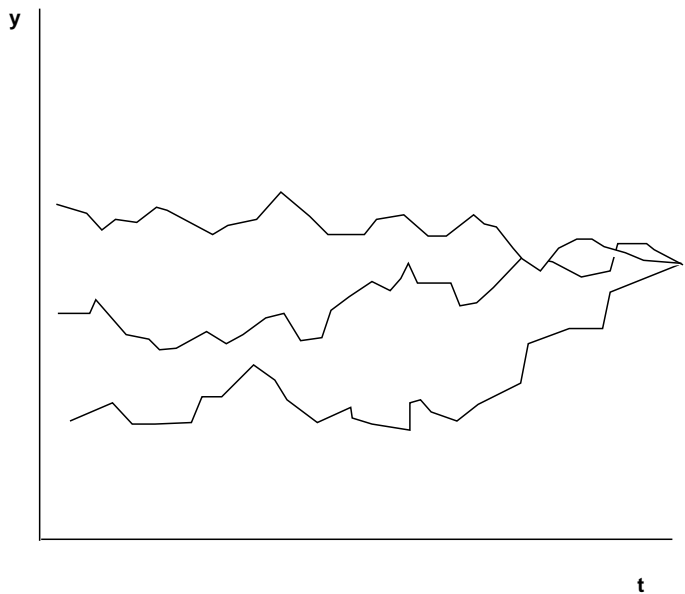}
}
\caption{Directed polymers in random media: The partition function (\ref{pathint})
is given by the sum over all directed paths $\vy(\tau)$ arriving at a given point,
each path being given a Boltzman weight depending on its elastic energy and the 
random pinning energy. The velocity in Burgers equation is proportional to the 
derivative of the polymer's free energy with respect to moving the arrival point
at fixed time}
\label{DPRM}
\end{figure}

The key ingredient at the heart of the solution of this problem is
the existence of a non linear transformation which transforms it into a stochastic 
but linear problem, the "Hopf-Cole" transformation, which has been known for
a long time \cite{Burgers}.
Define $Z(\vx,t)$ by $\v=-2 \nu \vec \nabla \log Z(\vx,t)$. It obeys the equation:
\be
{\partial Z \over \partial t} = \nu \nabla^2 Z + {\phi \over 2 \nu} Z
\label{poldir}
\ee
This is a Schr\"odinger equation in imaginary time, for a particle moving
in a time dependent random potential $\phi(\vx,t)/(2 \nu)$. It is
natural to introduce the corresponding (Euclidean) path integral
representation:

\be 
Z(\vx,t)=\int d \vy_0 \rho(\vy_0) \int_{\vy(0)=\vy_0}^{\vy(t)=\vx} {\cal D}(\vy)
\exp\((-{1 \over 2 \nu} \int_0^t d \tau \ \(({1 \over 2} \[[{\partial \vy \over \partial \tau}
\]]^2
+\phi(\tau,\vy(\tau)) \)) \))
\label{pathint}
\ee
where the integral ${\cal D}(\vy)$ is on all the paths $\vy(\tau)$ arriving
at $x$ for $\tau=t$, and the distribution  $\rho(\vy_0)$ encodes the
initial velocity field (see Fig.\ref{DPRM}). In this form one can identify the partition
function for a directed polymer in a random medium (DPRM). The
directed  polymer is characterised by the curve $\vy(\tau)$. It lives an a
$d+1$ dimensional space, which is the space-time of our original problem,
and it is directed in the 'time' direction. Its energy
is the sum of an elastic term, $E_{el}=\int d \tau \ 
\(({\partial \vy \over \partial \tau}\))^2$,
and a potential energy $E_p=\int d \tau \ \phi(\tau,\vy(\tau))$. This
potential energy corresponds to a random pinning potential of the polymer.
Identifying the temperature as $T={2\nu}$, we recognize in (\ref{pathint})
that each configuration $\vy(\tau)$ of the polymer is given a Boltzmann weight
$\exp-\(({E_{el}+E_p \over T} \))$. The function  $Z(\vx,t)$ is the
partition function for the polymers constrained
to arrive on the point $\vx,t$. In order to get back to the Burgers velocity
field, one must first compute the polymer's free energy $F(\vx,t)=- 2 \nu
\log Z(\vx,t)$ and take its gradient:
\be
\v(\vx,t) = - 2 \nu \vec \nabla \log Z(\vx,t) 
\label{hopfcole}
\ee
It is important to recognise that in the polymer language the potential
$\phi(\vx,t)$ is a {\it quenched} random potential: we need to evaluate
the free energy of the polymer in a fixed realisation of $\phi$ (a fixed sample),
and then do the sample averaging. This is not intuitive: In the original
problem $\phi$ is a time dependent field from which the stirring 
force derives. After the mapping it has to be considered as a quenched potential. 
This is because we have gone to a space-time description: if
we consider the initial Burgers problem in dimension $d$,
the space where
the directed polymer lives is $d+1$ dimensional. 

Directed polymers in random media have received a lot of attention
in recent years, both for their own interest and also for
their relevance in the description of magnetic flux tubes in type II
superconductors \cite{DPRM}. However these systems have a 
very important difference with respect to our problem issued from
Burger's turbulence: in general the correlation of the random
pinning potential are short range, and one is interested in what happens
on length scales much larger than this correlation. In our
problem we need to study the case where $\Delta$ is large, in the sense
that the inertial regime takes place on length scales much shorter than $\Delta$.
This will prevent us to use directly the known results on the DPRM. 

Yet
let us mention briefly some of the properties which are known for
the case of short range correlated disorder (this case can be studied also with a lattice
discretization of the space-time, each point of the lattice being given
an independent value of the potential).
 In this case, the statistical properties of DPRM also obey
some scaling relations, and no sign of intermittency has been reported so far.
 For instance the lateral wandering
of the polymer, far away from its extremities,
 is characterized by an exponent $\zeta$ defined
by:
\be
< E_T \[[\vy(\tau)-\vy(\tau'))^2 \]]> = C |\tau-\tau'|^{2 \zeta} \ .
\ee
where $E_T[A]$ is the expectation value  of the quantity $A$ with
the Boltzmann weight contained in (\ref{pathint}), while $<>$ stands as before
for the average over the random potential $\phi$.
Another exponent of interest is the exponent $\omega$
 characterizing the fluctuations
of the free energy when one varies the arrival point of the polymer:
\be
<(F(\vx,t)-F(\vx',t))^2> = C |\vx-\vx'|^{2 \omega/\zeta}
\ee
In order to get the velocity correlation function we need to differentiate this twice,
and we thus get:
$\alpha= \omega/\zeta-1$.
An important identity between the exponents is derived from Galilean invariance which
imposes that the terms ${\partial \v \over \partial t}$  and
$ (\v \cdot \vec \nabla) \ \v$ must have the same scaling behaviour, thus fixing \cite{galil}:
$\omega= 2 \zeta -1$. The exponents are known exactly
 only in $1+1$ space-time dimension, where $\zeta=2/3$. The general problem of the pinning
of elastic manifolds (of arbitrary internal dimension and codimension) by random impurities
is a major one in the statistical physics of disordered systems, but only
approximate answers are known so far \cite{DPRM,MP}.

We now close this parenthesis and get back to our problem where the
random potential is correlated on a length scale $\Delta$ which is much larger
than the (inertial) scale on which we want to study the system. A priori, this is a much simpler
problem: the random potential varies slowly, and thus we should be allowed to substitute
it by a linear potential. This is the essence of the random force approximation,
pioneered in the study of the pinning of elastic manifolds by Larkin and his
collaborators in the seventies 
\cite{larkin}:
\be
\phi(\tau,\vx)\simeq A(\tau)-\vec f(\tau). \vx(\tau)
\ee
In this case the total energy of the polymer is a quadratic function of the position, and
one can solve the problem completely. Introducing the Fourier transforms $\vx(\omega)$
and $\vf(\omega)$ of the fields $\vx(\tau)$ and $\vf(\tau)$, we find immediately the thermal
average  of the polymer's position $\vx$ (computed with the Boltzmann weight 
$(1/Z) \ \exp \[[-\int d \tau (\partial \vx / \partial \tau)^2+\int d \tau
\vf(\tau).\vx(\tau)\]]$):
\be
E_T(\vx(\omega))= {\vf(\omega) \over \omega^2}
\ee
Using the fact that $<\vf(\omega).\vf(\omega')> \propto \delta(\omega+\omega')$, one immediately
gets:
\be
<E_T\(( [\vx(\tau)-\vx(\tau')]^2 \))> \propto \int {d\omega \over \omega^4} [1-\cos(\omega (\tau-\tau'))]
\label{BuK41}
\ee
which seems to imply that the wandering exponent is $\zeta=3/2$. This is a very interesting
result since it leads to $\omega=2$ and $\alpha=1/3$. At first sight, we have derived $K41$
scaling from a two lines computation on the directed polymer, using only the
random force approximation which should be valid in the inertial range. However this is wrong.
The technical reason is obvious: the integral (\ref{BuK41}) over the 
frequency is divergent at small $\omega$. The consequences are far reaching:

\ni
- The divergence means that it is impossible to decouple the small scales from the large ones:
the random force approximation is never valid.

\ni
- We must thus study a DPRM in the presence of a full random potential, not
in a random force field. A crucial difference is that in presence of a random
potential there are many metastable states, while there is no metastability in
the random force approximation.

\ni
- As we shall see, the existence of metastable states of the potential is intimately related
to the breakdown of simple scaling, and the appearance of intermittency.

\section{The solution of forced Burgers turbulence in large dimensions}
So we face the difficult problem of computing the properties of a DPRM in the presence
of a random pinning {\it potential}, correlated on a scale $\Delta$. This may not look much easier than our
original Burgers problem, but it turns out that some of the powerful methods developped
in the study of disordered
systems 
 can be brought to bear on the DPRM. Technically this becomes rather involved, I shall just 
give the main steps of the computation.

The first step uses the replica method \cite{MPV}. One introduces the partition function $Z^n$
for $n$ replicas of the directed polymer, and averages it over the distribution of the random
potential. The analytic continuation to $n \to 0$ will thus provide the average free energy,
which we seek. The average velocity correlations will be obtained by a similar procedure
\cite{BMP}. The average of $Z^n$ is:
\be
<Z^n >= \prod_{a=1}^n \int {\cal D} (\vy_a) e^{-H_n}
\ee
where the average over disorder has introduced an effective {\it attraction} between
the various lines:
\be
{ H}_n =  \int_0^t d\tau   
\left[ {c \over 2} \sum_{a=1}^n  \(( {d \vy_a \over d\tau} \))^2 +
+ \sum_{a,b=1}^n C(\vy_a(\tau)-\vy_b(\tau)) \right]
\label{Hn}
\ee
where $C$ is the correlation function of the random pinning potential. In this language
we must study either the statistical mechanics of $n$ lines or the Euclidean quantum mechanics of $n$
particles, interacting by pairs through the potential $C$ (which is attractive and has a range 
equal to $\Delta$). We should compute the Green function for this problem and
then continue it analytically for $n \to 0$. So far this can be done exactly only in
two cases: the case of one dimension and $\Delta=0$ (giving back the exponent $\zeta=2/3$ of
usual DPRMs), or the infinite dimensional case, in which the variational method
developped below becomes exact.

To study this problem of $n$ interacting elastic lines, we have used a 
variational approach. It finds the best approximation to the system
by a quadratic Hamiltonian, belonging to the family:
\be
H_v= {1 \over 2} \sum_{a,b} \int d\tau \int d \tau' K_{ab}(\tau-\tau') \vx_a(\tau).\vx_b(\tau')
\ee
The idea is to compute the variational free energy, which is a functional
of $K_{ab} (\tau)$, and then to find the best set of functions $K_{ab} (\tau)$ which optimizes
the free energy. This variational method is a Hartree like approximation, which becomes
exact in the limit of large dimensions. It has been used a lot for the pinning of random manifolds
\cite{MP}, 
where it is able to give information on the phase diagram and the exponents in the 
glassy phase (the phase with many metastable states).
A priori, the problem is symmetric under the permutation of the replica indices. What is
found here is that, if the Reynolds number of the initial Burgers problem is
large enough (i.e. the pinning potential is
strong enough), there is a spontaneous breakdown of this permutation symmetry: In
the optimal solution, the various functions $K_{ab}, \ a\neq b$, are not all equal.
This phenomenon, called replica symmetry breaking (RSB), has been studied at length in
the spin glass problem \cite{MPV}. Finding a consistent scheme for RSB has been a major
achievement in the mean field model of spin glasses \cite{parisi}. Here we have used the same scheme
of RSB in our directed polymer problem. We shall not describe it
since it would take us too far (see \cite{BMP}). Instead we shall state the results 
of the variational method with RSB in physical terms. Intuitively, it is
enough to think of the RSB effect as being associated with the existence of many metastable
states: then ``different replicas may fall into different states''...

The physical description of the RSB solution is best described by the procedure
which generates a velocity pattern $\v(\vx,t)$ at a fixed time $t$, for one
given realization of the forcing. This procedure is as follows:

\noindent
- One chooses $M (>>1)$ points $\vr_\alpha$ independently, with a uniform distribution in
the box.

\noindent
- For each point, one chooses one "free energy" $f_\alpha$. These free energies are
independent, taken from
a Poisson process of density $\exp(f)$.

\begin{figure}
\centerline{\hbox{
\epsfig{figure=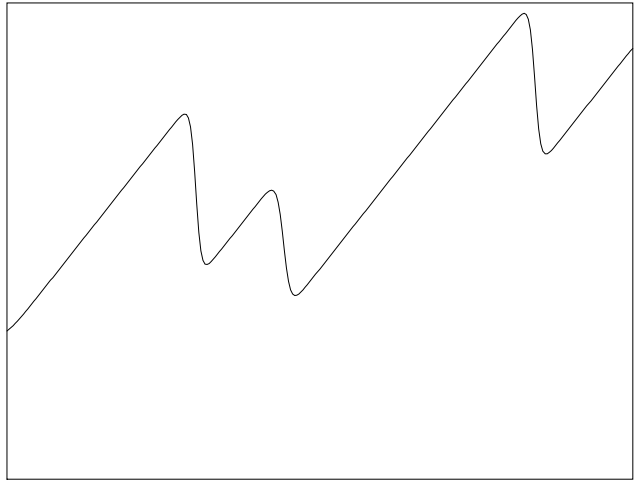,width=8cm,angle=-90}}
}
\caption{Schematic structure of Burgers flow in one dimension, obtained
by the replica variational approach to the DPRM formulation (\ref{Zres}).
The picture gives the velocity $v(x,t)$ as a function of $x$ for a given time $t$.
The typical width of the cells is of the order of the injection length scale $\Delta$,
while the width of the shocks behaves as $\Delta/Re$.
The structure is similar to that of decaying Burgers turbulence, and 
leads to an extreme type of intermittency effect(\ref{pdeu},\ref{expofin}).}
\label{1Dflow}
\end{figure}

One then generates the partition function as a weighted sum of
Gaussians:
\be
Z(\vx)=\sum_\alpha \exp\((-Re \[[ f_\alpha +{(\vx-\vr_\alpha)^2 \over 2 \Delta^2}\]]\))
\label{Zres}
\ee
and the velocity is  given by the Hopf Cole transformation (\ref{hopfcole}). The
number of states is irrelevant, as long as it is large, because of the
exponential distribution of the free energies, which ensures that the distribution
of the gaps (differences between the smaller free energies) is $M$ independent.
For large Reynolds numbers, the sum over the various points $\alpha$
is dominated, for each value of $\vx$, by one $\alpha$ which we call
$\alpha^*(\vx)$. This dominant $\alpha^*(\vx)$ is locally independent of $\vx$
(when $Re \to \infty$), but it jumps from time to time when
one varies $\vx$. Therefore, the structure of the flow
is locally radial:
\be
\v(\vx)=v_\Delta {\vx-\vr_{\alpha^*(\vx)} \over \Delta}
\ee
The flow organises itself into cells of typical size of order 
the injection scale $\Delta$, with a discontinuity of $\v$ (a 'shock')
between cells. From the representation (\ref{Zres}) one immediately sees that,
for finite $Re$, the region of the shock has a size of order $\Delta/Re$.
A typical structure of the flow in
one dimension is shown in Fig.\ref{1Dflow}. In higher dimensions, the 
regions of shocks separating the cells where the flow is radial have dimension $d-1$.

From this physical description of the replica solution one can infer the
scaling exponents. For simplicity the discussion is presented
in the $d=1$ case. Consider the velocity difference at distance $r$,
$u=|v( x+ r,t)-v( x,t)|$, in the inertial range $\Delta/Re
<<r<<\Delta$. 

\noindent
- If there is no shock in 
the range $[x,x+r]$, then $u=v_\Delta r/\Delta$.

\noindent
- If there is a shock, then $u$ is a random variable of typical
size of order $v_\Delta$. As the probability of there being a shock is of
order $r/\Delta$, one expects that the pdf of $u$ will take the following form:

\be
P(u)\sim (1-{r \over \Delta}) \delta\((u-v_\Delta {r \over \Delta}\)) +{r \over \Delta} f(u)
\label{pdeu}
\ee

This is indeed confirmed by the detailed replica computation, which
leads, in the limit of large $Re$,
to a complicated expression \cite{BMP,extremes}, the structure of which indeed
has the form (\ref{pdeu}). 
 Let us now estimate the various
moments of $u$ using expression (\ref{pdeu}): 
\be
\overline{u^p} \sim (1-{r \over \Delta})
({v_\Delta r \ov \Delta})^p + {r \over \Delta} \ A_p  \ ,
\ee
where $A_p =
\int du \ u^p f(u)$. In the inertial range, one finds that
for {\it all} $p$ larger than $1$, one has 
 \be
 \overline{u^p} = A_p
v_\Delta^p  {r \over \Delta} \ ,
\label{expofin}
 \ee
 whereas for $p<1$, one finds 
 \be
\overline{u^p} \propto v_\Delta^p  ({r \over \Delta})^p \ .
 \ee
So we face an extreme case of intermittency where all the exponents $\zeta_p$,
with $p>1$, are equal to one (see Fig.\ref{zetadep}). 
Notice that this intermittency has been found
here for stirred Burgers turbulence. The solution, which is based
on a variational method, is exact only in the limit of
large dimensionality. It is interesting to notice
however that the same spectrum of exponents $\zeta_p$ is found in
{\it decaying} Burgers turbulence in one dimension \cite{decay}, 
and has been called 
a `bifractal spectrum' (with reference to the multifractal
description of intermittency \cite{multifractal}). In fact there exists
a much deeper corrresondence, since our exact result for $P(u)$
in large dimensions and stirred turbulence is identical (up to rescalings
of lengths and velocities) is identical to the one obtained by Kida in
the decaying Burgers turbulence in $d=1$, in the case where the
initial velocity field is the gradient of a gaussian random
field with local correlations \cite{kida}. This is not fortuitous, and
is in fact due to the universality of extreme event statistics. This
universality, which leads to an exponential (`Gumbel')
distribution of low lying states, is probably at the heart of the success of the 
replica symmetry breaking in many situations, since the rsb is
well known to describe exactly such a distribution \cite{extremes,randfree}.

\section{The finite dimensional case}
A very natural question is: what survives of this large $d$ solution
in the finite dimensional case? Actually the stirred Burgers turbulence
in the finite $d$ case, and particularly
the case $d=1$, has been the subject of an intense activity in the recent years
\cite{yach,polyakov,gumi,e,slope}. The situation is still not totally
clarified, although interesting progress has been made. I shall not try to review
all the various methods which have been used, but just point out 
briefly a few facts in relation to our previous picture. 
At large Reynolds, the instantaneous velocity field
is a succession of regions where the velocity is continuous, separated
by shocks \cite{e}. The pdf of the velocity difference, $u$, has a `right
tail' (tail at positive $u$) which is dominated by the smooth regions. In these smooth
regions one can locally expand the velocity as
\be
v(x,t) \sim \lambda(t) (x-x_0(t))
\ee
and the slope $\lambda$ satisfies a Langevin type equation:
\be
{d \lambda \over dt} = -\lambda^2 + \eta
\ee
(where $\eta$ is a noise),
from which one deduces immediately \cite{slope} a right tail of the pdf of the slopes behaving
as $\exp(-C \lambda^3)$ and therefore a right tail of $P(u)$
behaving as $\exp(-C' u^3)$. This result was actually first found by Polyakov \cite{polyakov} using  a totally
different method: He writes the Hopf equation for the generating function
$<\exp\((\mu(v(x+r)-v(x))\))>$. This Hopf equation contains an anomalous term of the
type $\nu <v''(x) \exp\((\mu(v(x+r)-v(x))\))>$ which has to be dealt with,
and for which a certain form of operator product expansion has been proposed. 
This leads to the above right tail of $P(u)$. However 
the same right tail is found independently of the detailed form which is supposed for the 
operator product expansion, for which it does not provide a real test. 

The presence of this right tail in $\exp(-C' u^3)$ has been confirmed by several 
other approaches. One appealing formulation is the instanton approach of Gurarie
and Migdal\cite{gumi}. The procedure consists in writing the dynamical field theory
for the Burgers flow, and seeking the instanton configuration of the
velocity field, and the field conjugate to it, which provide the
tail of the pdf. In principle, such a procedure is allowed whenever one is interested in
some rare events. Here one fixes a value of $u$ which is improbably large, and one basically seeks the
most probable set of velocities and stirring which lead to this value. 
It can also be extended to higher dimensions where it leads to a similar 
right tail of $P(u)\sim \exp(-C' u^3)$, but the constant $C'$ turns
out to be proportional to the dimension $d$. This is the reason why such
a tail is not seen in the  (large $d$) solution of the previous sections.
 It also shows that the
large $u$ limit and the large $d$ limit do not always commute.

A much more complicated problem is the form of the pdf $P(u)$ at negative $u$.
The reason of the complication is clear: this negative $u$ region is  dominated
by shocks, and understanding it requires a control of the statistics of the shocks.
While this is doable in the case of decaying Burgers turbulence where the shocks
just merge, it is much more complicated in the forced case where new shocks keep appearing
at all times. So far the situation is not totally clear and some conflicting 
predictions have been made on the left tail, which I shall not try to review.

\section{Concluding remarks}
As mentioned above, I think that one  virtue of the large $d$ solution is
to provide an existence proof for an intermittency situation generated
by a non linear partial differential equation with a structure having
some similarities to that of Navier Stokes. Furthermore, the 
intermittency property is clearly associated with the existence of large scale
structures, the shocks, and it can be studied in many details.

To conclude I would like to point out some rather speculative idea which is another
interesting relation between disordered systems and Burgers turbulence, going in the
reverse direction. If one considers the pinning of elastic manifolds by a random potential,
there exist at the moment only two quantitative approaches to the study of this problem
starting from a microscopic description. One is the replica variational method
mentionned above \cite{MP}, the other one is the functional renormalization group
developped by D. Fisher and his collaborators \cite{frg}. It has been understood
recently that both approaches suggest the same picture for
 the effective free energy landscape of these manifolds at
large scale (at least for 
manifolds of internal dimension close to 4, embedded into
a space of large dimension): it is given by a succession of parabolic wells of random depth,
matching on singular points where the effective force is
discontinuous.  These parabolas are themselves subdivided into smaller
parabolas, corresponding to the motion of smaller length scales, in a
hierarchical manner \cite{babome}. Some preliminary investigation
\cite{boum} points towards the fact that
the renormalisation group flow for the gradient of the effective pinning potential is actually
described by a stirred Burgers-like equation (although the non linear term is superficially
irrelevant, it may become important because of the breakdown of scaling and the
generation of shocks). This would naturally account for the above picture, since
we know that a Burgers flow leads to a velocity pattern which is locally radial, 
associated thus with a locally parabolic effective pinning potential (free energy).
If it would be true, it would give another interesting perspective to the studies
of forced Burgers turbulence and the relevance of intermittency: the flow of the
effective pinning potential would get away from the gaussian subspace (which is
the only one studied so far), and one would rather need  to project it onto a different
subspace, that of locally parabolic functions, in order to get the correct 
fixed point and exponents.

\end{document}